# A Bayesian Approach to Approximate Joint Diagonalization of Square Matrices

**Mingjun Zhong**  MINGJUN.ZHONG@GMAIL.COM
Department of Biomedical Engineering
Dalian University of Technology, Dalian 116023, P.R.China

**Mark Girolami**  M.GIROLAMI@UCL.AC.UK
Department of Statistical Science
Centre for Computational Statistics and Machine Learning
University College London, London, WC1E 7HB, UK

## Abstract

We present a Bayesian scheme for the approximate diagonalisation of several square matrices which are not necessarily symmetric. A Gibbs sampler is derived to simulate samples of the common eigenvectors and the eigenvalues for these matrices. Several synthetic examples are used to illustrate the performance of the proposed Gibbs sampler and we then provide comparisons to several other joint diagonalization algorithms, which shows that the Gibbs sampler achieves the state-of-the-art performance on the examples considered. As a byproduct, the output of the Gibbs sampler could be used to estimate the log marginal likelihood, however we employ the approximation based on the Bayesian information criterion (BIC) which in the synthetic examples considered correctly located the number of common eigenvectors. We then succesfully applied the sampler to the source separation problem as well as the common principal component analysis and the common spatial pattern analysis problems.

## 1. Introduction

Joint diagonalization is a technique to simultaneously diagonalize a series of $K$ square symmetric matrices $\mathbf{C} = \{\mathbf{C}_k\}_{k=1}^K$ where $\mathbf{C}_k \in \mathcal{R}^{N \times N}$ (Friedland, 1983; Cardoso & Souloumiac, 1996; Pham, 2001;

van der Veen, 2001; Yeredor, 2002; Ziehe et al., 2004; Souloumiac, 2009). The standard joint diagonalization scheme finds a common matrix $\mathbf{W} \in \mathcal{R}^{N \times N}$ such that $\mathbf{W}\mathbf{C}_k\mathbf{W}^T$ for all $k$ are diagonal matrices. There are various applications for the joint diagonalization technique, such as blind source separation (BSS) (Belouchrani et al., 1997; Yeredor, 2002; Ziehe et al., 2004), common principal component analysis (CPCA) (Flury, 1984), and common spatial pattern analysis (CSPA) (Koles, 1991; Blankertz et al., 2008). However, it is well known that for more than two matrices it may not be possible to achieve joint diagonalization exactly and one must resort to approximate joint diagonalization for such applications. Furthermore, for some applications such as BSS, the matrices $\mathbf{C}_k$ may not even be symmetric.

In this paper, we generalize the standard joint diagonalization scheme by proposing an underlying statistical model. Rather than directly seeking $\mathbf{W}$, we infer a global matrix $\mathbf{B} \in \mathcal{R}^{N \times M}$, which is not necessarily square, and a series of $K$ diagonal matrices $\mathbf{\Lambda} = \{\mathbf{\Lambda}_k\}_{k=1}^K$ where $\mathbf{\Lambda}_k = diag(\lambda_1^k, \lambda_2^k, \cdots, \lambda_M^k)$ are the eigenvalues for every $k$

$$\mathbf{C}_k = \mathbf{B}\mathbf{\Lambda}_k\mathbf{B}^T + \mathbf{E}_k \qquad (1)$$

where $\mathbf{E}_k = (\mathbf{e}_1^k, \mathbf{e}_2^k, \cdots, \mathbf{e}_N^k)$ are assumed errors. A Gaussian error model $\mathbf{e}_n^k \sim \mathcal{N}(0, \sigma_k^2\mathbf{I})$ is assumed, therefore each element $e_{ij}^k$ of $\mathbf{E}_k$ follows a Gaussian distribution with zero mean and variance $\sigma_k^2$. We note that in this formulation the matrices $\mathbf{C}_k$ may not be symmetric. In this paper we restrict ourselves to the case $\mathbf{B}^T\mathbf{B} = \mathbf{I}$, however further work could consider the case where $\mathbf{B}$ is non-orthogonal (Yeredor, 2002; Ziehe et al., 2004; Souloumiac, 2009). The columns of $\mathbf{B}$ are then recognized as the common eigenvectors.

This model is a generalization of the subspace fitting models of (van der Veen, 2001; Yeredor, 2002), where here the errors $\mathbf{E}_k$ are represented explicitly.

Various joint diagonalization algorithms have been proposed in the literature. The Jacobi method of (Cardoso & Souloumiac, 1996) seeks an orthogonal matrix $\mathbf{W}$ for jointly diagonalizing $\mathbf{C}_k$ whilst Pham's algorithm (Pham, 2001) is restricted to positive definite matrices $\mathbf{C}_k$. Compared to these methods, our method is not restricted to positive definite $\mathbf{C}_k$ furthermore $\mathbf{B}$ could be a non-square matrix. Our approach is most similar to the subspace fitting models (van der Veen, 2001; Yeredor, 2002) although we also treat the error variances as parameters to be inferred. Flurry (Flury, 1984) and Hoff (Hoff, 2009b) proposed CPCA for a series of covariance matrices whose eigenvalues were assumed positive, a restriction which we relax.

Point estimate methods such as the expectation-maximisation (EM) algorithm (Dempster et al., 1977) or the AC-DC algorithm (Yeredor, 2002) could be derived to estimate these parameters. In this paper we prefer a Bayesian formulation for the model (1) where the posterior distributions for parameters is provided. A Gibbs sampler is derived to simulate model parameters from their posterior distribution. Based on some data, we compared the Gibbs sampler to the Jacobi method (Cardoso & Souloumiac, 1996), AC-DC (Yeredor, 2002) and the FFDiag algorithm (Ziehe et al., 2004). Across all the results the Gibbs sampler is comparable to both the Jacobi and FFDiag methods whilst outperforming the AC-DC algorithm in terms of the Amari performance index (Amari et al., 1996).

## 2. The Model

### 2.1. The Data Likelihood

Given the model in (1) and the assumed error distribution the likelihood, $p(\mathbf{C}|\Theta)$, has the following form

$$\prod_{k=1}^{K} |2\pi\sigma_k^2 \mathbf{I}|^{-\frac{1}{2}} \exp\left\{-\frac{1}{2\sigma_k^2}||\mathbf{C}_k - \mathbf{B}\mathbf{\Lambda}_k\mathbf{B}^T||_F^2\right\}$$

where $\Theta = (\mathbf{B}, \mathbf{\Lambda}, \sigma)$ and $||X||_F^2 = trace(X^T X)$ is the Frobenius norm. For simplicity of representing the conditional distributions, we give an alternative representation of the model (1). Let $vec(\cdot)$ denote the transformation of a matrix to a vector by concatenation of the columns. Denote $\mathbf{x}_k = vec(\mathbf{C}_k)$,

$\mathbf{A} = (\mathbf{B} \otimes \mathbf{1}) \odot (\mathbf{1} \otimes \mathbf{B})$ where $\mathbf{1}$ denotes a column vector of size $N \times 1$ with all the elements being 1, $\mathbf{u}_k = (\lambda_1^k, \lambda_2^k, \cdots, \lambda_M^k)^T$ and $\epsilon_k = vec(\mathbf{E}_k)$. Then the model (1) could alternatively be represented as $\mathbf{x}_k = \mathbf{A}\mathbf{u}_k + \epsilon_k$. We denote the data $\mathbf{X} = (\mathbf{x}_1, \mathbf{x}_2, \cdots, \mathbf{x}_K)$ with dimension $N^2 \times K$ and likelihood accordingly

$$p(\mathbf{X}|\Theta) = \prod_{k=1}^{K} |2\pi\sigma_k^2 \mathbf{I}_{N^2}|^{-\frac{1}{2}} \exp\left\{-\frac{1}{2\sigma_k^2}||\mathbf{x}_k - \mathbf{A}\mathbf{u}_k||_2^2\right\}$$

Given the data likelihood and associated priors on the parameters the posterior distribution takes the form

$$p(\mathbf{B}, \mathbf{U}, \sigma|\mathbf{X}) \propto p(\mathbf{X}|\mathbf{B}, \mathbf{U}, \sigma)p(\mathbf{B})p(\mathbf{U})p(\sigma)$$

which is detailed in the following subsection.

### 2.2. The Priors and Posteriors

We employ a Gaussian prior on $\mathbf{u}_k$ depending on $\sigma_k^2$ and a hyperparameter $v_k^2$ such that

$$p(\mathbf{u}_k|\sigma_k^2, v_k^2) = |2\pi\sigma_k^2 v_k^2 \mathbf{I}_M|^{-\frac{1}{2}} \exp\left\{-\frac{1}{2\sigma_k^2 v_k^2}\mathbf{u}_k^T\mathbf{u}_k\right\}$$

The conditional posterior for $\mathbf{u}_k$ has the exact form

$$p(\mathbf{u}_k|\mathbf{x}_k, \mathbf{A}, \sigma_k, v_k) \sim \mathcal{N}(\mu_{\mathbf{u}_k}, \sigma_k^2 \mathbf{\Sigma}_{\mathbf{u}_k})$$

where $\mathbf{\Sigma}_{\mathbf{u}_k} = (v_k^{-2}\mathbf{I}_M + \mathbf{A}^T\mathbf{A})^{-1}$ & $\mu_{\mathbf{u}_k} = \mathbf{\Sigma}_{\mathbf{u}_k}\mathbf{A}^T\mathbf{x}_k$.

An inverse Gamma distribution is employed for $\sigma_k^2$ such that $p(\sigma_k^2|a_k, b_k) \propto (\sigma_k^2)^{-a_k-1}\exp\{-b_k\sigma_k^{-2}\}$, thus the conditional posterior is

$$p(\sigma_k^2|\mathbf{x}_k, \mathbf{A}, \mathbf{u}_k, v_k) \sim InvGamma(a_{\sigma_k^2}, b_{\sigma_k^2}) \quad (2)$$

where $a_{\sigma_k^2} = a_k + \frac{1}{2}N^2 + \frac{1}{2}M$ and $b_{\sigma_k^2} = b_k + \frac{1}{2}||\mathbf{x}_k - \mathbf{A}\mathbf{u}_k||_2^2 + \frac{1}{2}v_k^{-2}\mathbf{u}_k^T\mathbf{u}_k$. So $p(\sigma_k^{-2}|\mathbf{x}_k, \mathbf{A}, \mathbf{u}_k, v_k) \sim Gamma(a_{\sigma_k^2}, 1/b_{\sigma_k^2})$ is a Gamma distribution. We also require an inverse Gamma prior for $v_k^2$ such that $p(v_k^2|a_k, b_k) \sim InvGamma(a_k, b_k)$, so the posterior for $v_k^2$ is

$$p(v_k^2|\mathbf{u}_k, \sigma_k^2) \sim InvGamma(a_{v_k^2}, b_{v_k^2})$$

where $a_{v_k^2} = a_k + \frac{1}{2}M$ and $b_{v_k^2} = b_k + \frac{1}{2}\sigma_k^{-2}\mathbf{u}_k^T\mathbf{u}_k$. So $p(v_k^{-2}|\mathbf{u}_k, \sigma_k^2) \sim Gamma(a_{v_k^2}, 1/b_{v_k^2})$.

To derive the conditional posterior for $\mathbf{B}$, it is convenient to employ the Frobenius norm form of the data likelihood. Since $\mathbf{B}^T\mathbf{B} = \mathbf{I}$, we assume a uniform distribution on the Stiefel manifold for the random matrix

**Algorithm 1** Gibbs Sampler
  **Input:** Matrices $\mathbf{C}_k$ ($k = 1, \cdots, K$), number of eigenvectors $M$, number of samples $NSamps$
  Initialize $\mathbf{B}$, $\mathbf{U}$, $\sigma_k^{-2}$, $v_k^{-2}$
  **for** $j = 1$ **to** $NSamps$ **do**
    **for** $k = 1$ **to** $K$ **do**
      (a) Sample $\mathbf{u}_k \sim \mathcal{N}(\mu_{\mathbf{u}_k}, \sigma_k^2 \mathbf{\Sigma}_{\mathbf{u_k}})$
      (b) Sample $\sigma_k^{-2} \sim Gamma(a_{\sigma_k^2}, 1/b_{\sigma_k^2})$
      (c) Sample $v_k^{-2} \sim Gamma(a_{v_k^2}, 1/b_{v_k^2})$
    **end for**
    Sample $\mathbf{B} \sim \mathcal{LB}(\Lambda, \Upsilon)$
  **end for**

$\mathbf{B}$ (Chikuse, 2003). The posterior for $\mathbf{B}$ then has the following form,

$$p(\mathbf{B}|\mathbf{C}, \Lambda, \sigma) \propto \prod_{k=1}^{K} \exp\left\{tr\left(\Lambda_k \mathbf{B}^T \frac{\mathbf{C}_k^T + \mathbf{C}_k}{2\sigma_k^2} \mathbf{B}\right)\right\}$$

$$= \prod_{m=1}^{M} \exp\left\{\mathbf{b}_m^T \sum_{k=1}^{K} \frac{\lambda_m^k(\mathbf{C}_k^T + \mathbf{C}_k)}{2\sigma_k^2} \mathbf{b}_m\right\}$$

This is the matrix Langevin-Bingham (LB) distribution or the matrix Bingham-von Mises-Fisher distribution (Khatri & Mardia, 1977). For simplicity, denote $\Upsilon_k = \frac{\mathbf{C}_k^T + \mathbf{C}_k}{2\sigma_k^2}$ and $\Upsilon = \{\Upsilon_k\}_{k=1}^K$. We denote the matrix LB distribution as $\mathcal{LB}(\Lambda, \Upsilon)$. It should be noted that even though the distribution is represented as a product form, since $\mathbf{B}^T\mathbf{B} = \mathbf{I}$, the columns of $\mathbf{B}$ are not statistical independent. Given those conditional posterior distributions, a Gibbs sampler in Algorithm 1 is derived for simulating the parameters. We require to simulate the matrix LB distribution which is described in the following section.

## 3. Sampling from the matrix Langevin-Bingham distribution

### 3.1. The vector Bingham distribution

To sample from the matrix LB distribution, we need to draw samples from the following vector Bingham distribution

$$p(\mathbf{x}|\Sigma) = c^{-1}(\Sigma) \exp(\mathbf{x}^T \Sigma \mathbf{x})$$

where $\Sigma$ is symmetric, and $\mathbf{x} \in \mathcal{R}^M$ such that $\mathbf{x}^T\mathbf{x} = 1$. This distribution is defined under the uniform measure $d_{\mathcal{S}^{M-1}}(\mathbf{x})$ in the sphere $\mathcal{S}^{M-1}$. The uniform measure $d_{\mathcal{S}^{M-1}}(\mathbf{x})$ is invariant under the orthogonal transformation. Therefore, for any orthogonal matrix $\Gamma \in \mathcal{O}(M)$, set $\mathbf{y} = \Gamma^T\mathbf{x}$, then $\mathbf{y}$ has a Bingham distribution $p(\mathbf{y}|\Gamma^T\Sigma\Gamma)$ with respect to the uniform measure in $\mathcal{S}^{M-1}$. The eigenvalue decomposition could be applied to $\Sigma$, then $\Sigma = U\Lambda U^T$, where $\Lambda = diag(\lambda_1, \lambda_2, \cdots, \lambda_M)$. Set $\mathbf{y} = U^T\mathbf{x}$, then $\mathbf{y}$ has a simplified Bingham distribution $p(\mathbf{y}|\Lambda)$. Instead of sampling $p(\mathbf{x}|\Sigma)$ directly, we sample the density of $\mathbf{y}$

$$p(\mathbf{y}|\Lambda) = c^{-1}(\Lambda) \exp(\mathbf{y}^T \Lambda \mathbf{y})$$

with respect to the uniform measure in $\mathcal{S}^{M-1}$ such that $\sum_{i=1}^M y_i^2 = 1$. The uniform measure in $\mathcal{S}^{M-1}$ has the following form with respect to the Lebesgue measure in $\mathcal{R}^M$ (Kume & Walker, 2006),

$$d_{\mathcal{S}^{M-1}}(\mathbf{y}) \propto \left(1 - \sum_{i=1}^{M-1} y_i^2\right)^{-\frac{1}{2}} \mathbf{1}\left(\sum_{i=1}^{M-1} y_i^2 < 1\right) \prod_{i=1}^{M-1} dy_i$$

The random vector $\mathbf{y}$ on $\mathcal{S}^{M-1}$ has the following well known representations (Chikuse, 2003)

$$y_1 = \cos\theta_1$$
$$y_2 = \sin\theta_1 \cos\theta_2$$
$$\cdots\cdots$$
$$y_{M-1} = \sin\theta_1 \sin\theta_2 \cdots \sin\theta_{M-2} \cos\theta_{M-1}$$
$$y_M = \sin\theta_1 \sin\theta_2 \cdots \sin\theta_{M-2} \sin\theta_{M-1}$$

Instead of directly drawing samples for $y_1, \cdots, y_{M-1}$ we could draw samples for $\theta_1, \cdots, \theta_{M-1}$. However, for simplicity, given $\theta = \cos^2\theta_1$, we could employ a new parameterisation of the form $y_1^2 = \theta$ and $y_i^2 = (1-\theta)u_i$, where $i = 2, \cdots, M$, which induces the following Lebesgue measure

$$d_{\mathcal{S}^{M-1}}(\mathbf{y}) \propto s_1 \theta^{-\frac{1}{2}}(1-\theta)^{\frac{M-3}{2}} \left(1 - \sum_{i=2}^{M-1} u_i\right)^{-\frac{1}{2}}$$
$$\mathbf{1}\left(\sum_{i=1}^{M-1} y_i^2 < 1\right) d\theta \prod_{i=2}^{M-1} s_i u_i^{-\frac{1}{2}} du_i$$

where $s_i \in \{-1, +1\}$ is the sign of the variable $y_i$. Consequently, the vector Bingham density has the following form with respect to $\theta, u_i$ where $i = 2, \cdots, M$

$$p(\theta, u_i|\Lambda) \propto \exp\left\{\lambda_1 \theta + (1-\theta) \sum_{i=2}^{M} \lambda_i u_i^2\right\}$$
$$\times s_1 \theta^{-\frac{1}{2}}(1-\theta)^{\frac{M-3}{2}} \left(1 - \sum_{i=2}^{M-1} u_i\right)^{-\frac{1}{2}} \prod_{i=2}^{M-1} s_i u_i^{-\frac{1}{2}}$$

Conditional on $u_2, \cdots, u_{M-1}$, we draw samples for $\theta$. Given the current known values of $\mathbf{y}$, $u_i = \frac{y_i^2}{1-y_1^2}$ for $i = 2, \cdots, M$. We draw samples for $\theta$ proportional to $p(\theta) \propto \theta^{-\frac{1}{2}}(1-\theta)^{\frac{M-3}{2}} \exp\left\{\lambda_1\theta + (1-\theta)\sum_{i=2}^{M} \lambda_i u_i^2\right\}$.

We could employ a rejection scheme as in (Hoff, 2009a). Since this distribution has a simple form and $\theta \in (0,1)$, a grid or Slice sampling (Neal, 2003) could also be employed. For the $s_i$, it is just sampled uniformly from $\{-1, +1\}$. Given the sampled $\theta^*$, a new sample for $\mathbf{y}$ is then $y_1 = s_1\sqrt{\theta^*}$ and $y_i = s_i\sqrt{(1-\theta^*)u_i}$ for $i = 2, \cdots, M$. Finally a new sample $\mathbf{x}^* = U\mathbf{y}^*$.

### 3.2. The matrix Langevin-Bingham distribution

We employ the procedures of Hoff (2009a) to simulate the matrix LB distribution. When $M < N$, we denote $\mathbf{G}_m = \sum_{k=1}^{K} \frac{\lambda_m^k}{2\sigma_k^2}(\mathbf{C}_k^T + \mathbf{C}_k)$. We then need to sample the matrix LB distribution of the following form,

$$p(\mathbf{B}|\mathbf{C}, \Lambda, \sigma) \propto \prod_{m=1}^{M} \exp\left\{\mathbf{b}_m^T \mathbf{G}_m \mathbf{b}_m\right\}$$

Since the columns of $\mathbf{B}$ are orthogonal to each other, $\mathbf{b}_m$ are not statistically independent. To draw any $\mathbf{b}_i$, we denote the rest of the columns of $\mathbf{B}$ by $\mathbf{B}_{-i}$. Denote $\mathbf{Q}$ having the size $M \times (M-1)$ as the orthonormal basis for the Null space of $\mathbf{B}_{-i}$. Then we could have the representation $\mathbf{b}_i = \mathbf{Q}\beta$ for $\beta \in \mathcal{S}^{M-2}$. So the conditional distribution of $\beta$ given $\mathbf{B}_{-i}$ has the following representation

$$p(\beta|\mathbf{B}_{-i}, \mathbf{G}_m) \propto \exp\left\{\beta^T \widetilde{\mathbf{G}}_m \beta\right\}$$

where $\widetilde{\mathbf{G}}_m = \mathbf{Q}^T \mathbf{G}_m \mathbf{Q}$. Since $\beta^T \beta = 1$, it is a vector Bingham distribution. We can in the first step draw a sample $\beta^*$, and then the new sample for $\mathbf{b}_i^*$ is obtained by the transformation $\mathbf{Q}\beta^*$.

For $M = N$, we need to simulate two columns of $\mathbf{B}$ at the same time (Hoff, 2009a). Without loss of generality, we look at sampling the first two columns of $\mathbf{B}$. We denote $\mathbf{B}_{-(1,2)}$ as the matrix $\mathbf{B}$ without the first two columns. Let $\mathbf{Q}$ be the orthonormal basis for the Null space of $\mathbf{B}_{-(1,2)}$. Then $(\mathbf{b}_1, \mathbf{b}_2) = \mathbf{Q}Z$ for a proper $2 \times 2$ orthonormal matrix $Z = (z_{ij})$. Instead of sampling $(\mathbf{b}_1, \mathbf{b}_2)$ directly, we sample $Z$. Denote $\widetilde{A}_k = \mathbf{Q}^T \frac{1}{2\sigma_k^2}(\mathbf{C}_k^T + \mathbf{C}_k)\mathbf{Q}$ and $\widetilde{\Lambda}_{1,2}^k = diag(\lambda_1^k, \lambda_2^k)$. The posterior for $Z$ is

$$\begin{aligned} p(Z|\mathbf{B}_{-(1,2)}, \widetilde{A}_k, \widetilde{\Lambda}_{1,2}^k) &\propto \prod_{k=1}^{K} \exp\left\{tr\left(\widetilde{\Lambda}_{1,2}^k Z^T \widetilde{A}_k Z\right)\right\} \\ &= \exp\left\{Z_{.1}^T A_1 Z_{.1} + Z_{.2}^T A_2 Z_{.2}\right\} \end{aligned}$$

where $Z_{.i}$ denotes the $i$th column of $Z$ and $A_i = \sum_{k=1}^{K} \lambda_i^k \widetilde{A}_k$ where $i = 1, 2$. Given a sample $Z^*$, a new sample is then $(\mathbf{b}_1^*, \mathbf{b}_2^*) = \mathbf{Q}Z^*$.

## 4. Simulation Results

In this section we evaluate the proposed Gibbs sampler based on a few data sets. In the first example, based on a toy data set we compare the Rejection, Slice and Grid sampling schemes for simulating the vector Bingham distribution. In the second example, the Gibbs sampler was applied to data with size $N = 10$, $M = 5$ and $K = 100$ to seek the common eigenvectors. We also considered data with size $N = 10$, $M = 10$ and $K = 100$. Given these results, we compare the Gibbs sampler to the Jacobi method, AC-DC and FFDiag algorithms. We finally applied the Gibbs sampler to BSS, CPCA and CSPA. Note that all the Gibbs samplers in our experiments were monitored for convergence. For each sampler we used 10 chains to calculate the $R$ statistics defined in (Gelman & Rubin, 1992) and the Gibbs sampler was considered converged when $R < 1.2$. All the results shown here were the samples drawn after convergence to the invariant measure.

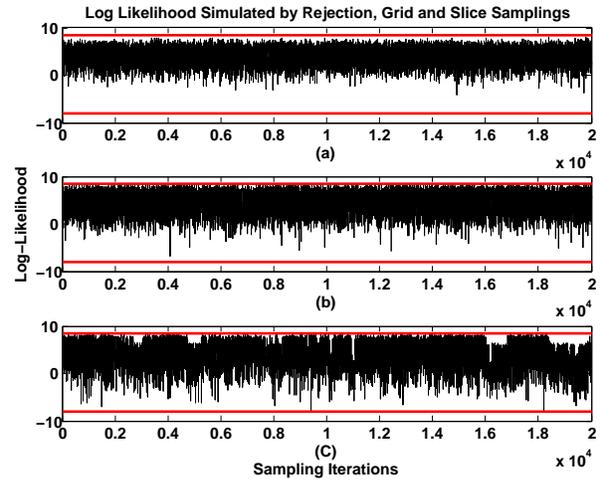

Figure 1. Log-Likelihood simulated by using the rejection (a), grid (b) and slice (c) sampling schemes. The minimum and maximum eigenvalues are denoted by the red lines.

### 4.1. Simulating vector Bingham distribution

In this experiment, we compare the three sampling schemes, namely rejection, slice and grid samplings, for simulating the vector Bingham distributions $p(\mathbf{x}|\mathbf{A}) \propto \exp(\mathbf{x}^T \mathbf{A} \mathbf{x})$. A symmetric real matrix $\mathbf{A}$ with size $10 \times 10$ was randomly generated which has 10 different eigenvalues. We calculated the effective sample sizes (ESS) for all the methods which are shown in Table 1. The simulations of the log likelihood for the three methods are plotted in Figure 1. As expected, all the values of log likelihood lie in between the minimum and maximum eigenvalues of $\mathbf{A}$. We conclude that all the three schemes perform similarly in terms of ESS. We then applied all the schemes to the Gibbs sampler for the joint diagonalization problems.

Table 1. ESS statistics for the rejection, slice and grid sampling schemes for the vector Bingham distribution.

| Schemes | Min | Median | Mean | Max |
|---|---|---|---|---|
| Rejection | 19214 | 19678 | 19682 | 20000 |
| Slice | 18703 | 19629 | 19550 | 20000 |
| Grid | 18960 | 19895 | 19722 | 20000 |

Table 2. ESS statistics for the rejection, slice and grid sampling schemes for the matrix LB distribution.

| Schemes | Min | Median | Mean | Max |
|---|---|---|---|---|
| Rejection | 2292 | 2578 | 2553 | 2758 |
| Slice | 3972 | 5358 | 5209 | 5917 |
| Grid | 116 | 170 | 179 | 279 |

### 4.2. Gibbs sampler for joint diagonalization

In this section we evaluate the proposed Gibbs sampler for the joint diagonaliztion problems based on two data sets. For the data considered, $K = 100$. The matrices were generated according to $\mathbf{C}_k = \mathbf{B}\Lambda_k\mathbf{B}^T + \mathbf{E}_k$ where $\mathbf{B}^T\mathbf{B} = \mathbf{I}$, and each matrix $\mathbf{C}_k$ has size $10 \times 10$. All the diagonal elements of the diagonal matrix $\Lambda_k$ were randomly generated. For the first data set, the matrix $\mathbf{B}$ has size $10 \times 5$. For the second data set, the matrix $\mathbf{B}$ has size $10 \times 10$. Since Gaussian errors were assumed the matrices $\mathbf{C}_k$ may not be symmetric. All the algorithms considered here were designed to seek $\mathbf{B}$ and $\Lambda_k$. If $\widehat{\mathbf{B}}$ is denoted as the estimation of $\mathbf{B}$, for comparison purposes we calculate Amari's performance index (API) (Amari et al., 1996) for the matrix $\mathbf{P} = \widehat{\mathbf{B}}^{-1}\mathbf{B}$. Note that when $\mathbf{P}$ is a permutation of the identity matrix, the API is zero. The API is defined as $\sum_i \left(\sum_j \frac{|P_{ij}|}{\max_l |P_{il}|} - 1\right) + \sum_j \left(\sum_i \frac{|P_{ij}|}{\max_l |P_{lj}|} - 1\right)$.

#### 4.2.1. The $10 \times 5$ matrix $\mathbf{B}$

For the first experiment when $M < N$, we compare the performance of the Gibbs sampler based on rejection, Slice and grid sampling schemes for sampling from the matrix LB distribution. We calculated the ESS for the three sampling schemes which are shown in the Table 2. We see that both the rejection and slice sampling are more efficient than the grid sampling in terms of ESS. We also plotted the log likelihood for the three schemes (see Figure 2), which shows that the grid sampling scheme performs better than the other two in searching for the modes of the model. We now employ the grid sampling scheme for the model when $N > M$. The ESS of grid sampling was very small, which suggests that grid sampling may be performing a random walk in the local mode of the posterior.

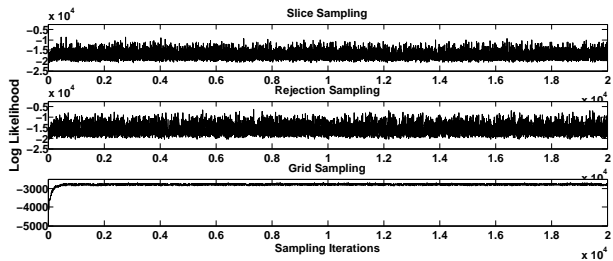

Figure 2. Log-Likelihood simulated by using the rejection, grid and slice samplings for the matrix LB distribution.

Since our model is similar to that of (Yeredor, 2002), we compare grid sampling to the AC-DC algorithm. Both the grid sampling and AC-DC algorithm were applied to the data to infer $\mathbf{B}$ and $\Lambda_k$. We calculated the API values for the estimates from both methods when various errors were added to the model. The API statistics for grid sampling and the minimum API values for the AC-DC algorithm are shown in the Table 3. Note that we also show the API computed by using the maximum a posteriori (MAP) sample. The results show that the proposed scheme outperforms the AC-DC algorithm across all error levels. We note here that one of the advantages of the Bayesian approach compared to point estimate methods is that the sampling approach accommodates naturally the model selection issue. As the first step to model selection, we used the Gibbs sampling output to approximate the log marginal likelihood of various models using the BIC (Schwarz, 1978). We calculated the log marginal likelihood for the models $M = 1, 2, \cdots, N$ where $M$ denotes the number of columns of $\mathbf{B}$. The estimates are shown in the Figure 3 which indicates that BIC prefers $M = 5$ which correctly located the model. Note that besides the approximate BIC method considered here more rigorous (i.e. unbiased estimation schemes) could be used for model selection as in (Zhong et al., 2011)

#### 4.2.2. The $10 \times 10$ matrix $\mathbf{B}$

For the second experiment, we applied grid sampling, AC-DC, Jacobi and FFDiag methods to the data set where the matrix $\mathbf{B}$ has size $10 \times 10$. The grid sampling scheme was used to sample the matrix $\mathbf{B}$ and all the rest algorithms were used to provide point estimates for $\mathbf{B}$. The API statistics for grid sampling are shown in the Table 4. The minimum API values for AC-DC, Jacobi method and FFDiag are shown in Table 5. These results show that the proposed method is comparable to the Jacobi method and outperforms both the AC-DC and FFDiag in terms of the minimum API. However, in practice we usually do not know the true $\mathbf{B}$. So the MAP sample could be used for comparison purposes, which shows that the Gibbs sampler is comparable to FFDiag and outperforms AC-DC.

Table 3. The API values of the Gibbs sampler and AC-DC.

| Noise Var. | Grid Sampling Min; Mean (std); Max; MAP | AC-DC |
|---|---|---|
| 0.01 | 0.1343;0.2508(0.0398);0.3521;0.1839 | 0.2435 |
| 0.1 | 0.2599;0.4961(0.0894);0.9124;0.4624 | 0.6040 |
| 0.5 | 0.6317;1.1259(0.1937);1.6440;0.8930 | 1.2844 |
| 1 | 1.3440;3.7004(0.8129);5.8577;3.1732 | 3.2071 |

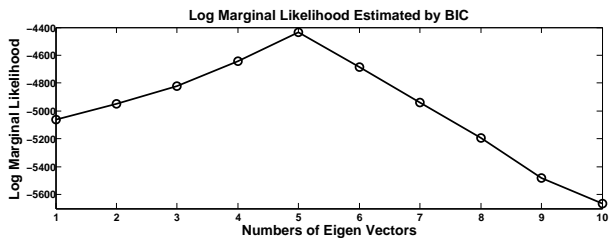

Figure 3. Log marginal likelihood estimated by BIC for models with various number of eigenvectors.

### 4.3. Applications to BSS

In this section we apply the proposed approximate joint diagonalization scheme to a BSS problem. We employ the 'ACsin10d' data from the ICAlab benchmark data sets (http:// www.bsp.brain.riken.jp/ ICALAB/). This data set contains 10 sine-wave sources denoted by $\mathbf{S}$, and each source is sampled 1000 equally spaced data points. We randomly generated a mixing matrix $\mathbf{A}$ of size $10 \times 10$ and then the mixtures were generated by $\mathbf{X} = \mathbf{AS} + \mathbf{E}$ where $\mathbf{E}$ were the Gaussian error realisations with $\sigma = 0.1$. The data $\mathbf{X}$ were whitened by multiplying the matrix $\mathbf{W}$ as defined in (Belouchrani et al., 1997) and then we obtain 100 covariance matrices $\mathbf{C}(\tau) = E\left\{\widetilde{\mathbf{X}}(t+\tau)\widetilde{\mathbf{X}}(t)^T\right\}$. For comparison purposes, we applied the Jacobi method, FFDiag algorithm, AC-DC algorithm as well as grid sampling of $\mathbf{C}(\tau)$ to search for the diagonalizing matrix $\mathbf{B}$. We then calculated the API values for the matrix $\mathbf{P} = \mathbf{B}^{-1}\mathbf{WA}$. The API values for the Jacobi method, FFDiag and AC-DC were 1.0333, 1.4951 and 1.5018. The API value statistics obtained from grid sampling were 1.1869 (Min), 1.3046±0.0361 (Mean±std) , 1.4408 (Max) and 1.3254 (MAP), indicating that the proposed method produced smaller API values than both the FFDiag and AC-DC with the Jacobi method always achieving the best API.

Table 4. API values computed by the grid sampling.

| Var. | Min | Mean (std) | Max | MAP |
|---|---|---|---|---|
| 0.01 | 0.0488 | 0.0727 (0.0075) | 0.1020 | 0.0548 |
| 0.1 | 0.1108 | 0.1658 (0.0163) | 0.2252 | 0.1501 |
| 0.5 | 0.2184 | 0.3277 (0.0350) | 0.4194 | 0.3316 |
| 1 | 0.3281 | 0.4990 (0.0536) | 0.6709 | 0.4808 |

Table 5. The minimum API values computed by Jacobi method, AC-DC algorithm and FFDiag algorithm.

| Noise Var. | Jacobi | AC-DC | FFDiag |
|---|---|---|---|
| 0.01 | 0.0322 | 0.0757 | 0.0477 |
| 0.1 | 0.1155 | 0.2574 | 0.1654 |
| 0.5 | 0.2081 | 0.4860 | 0.3008 |
| 1 | 0.3715 | 0.7878 | 0.5707 |

### 4.4. Applications to CPCA and CSPA

In this example, we show that the Gibbs sampler could be applied to both CPCA and CSPA. Given $K$ covariance matrices $\mathbf{C}_k$, CPCA seeks to find a common orthogonal matrix $\mathbf{B}$ such that $\mathbf{BC}_k\mathbf{B} = \Lambda_k$ are diagonal (Flury, 1984). The columns of $\mathbf{B}$ are the common principal components (CPC). However, for non-trivial problems it would be unlikely that the $\Lambda_k$ are exactly diagonal. We employ the model (1) with Gaussian errors to infer the CPC. CPCA can be applied to signal processing, such as the CSPA which has been applied to the preprocessing of EEG data in Brain Computer Interfaces (BCI) (Blankertz et al., 2008). We first apply the Gibbs sampler to a toy data set and then some EEG data.

#### 4.4.1. A Synthetic Data

We employ the procedure in (Blankertz et al., 2008) to generate data in demonstrating the Gibbs sampler in CPCA and CSPA. The data for analysis is generated according to the following linear mixing model

$$Y^j = AS^j$$

where $j = 1,2$ are class labels, and $S^1 \sim \mathcal{N}(0,\Lambda^1)$ where $\Lambda^1 = diag(0.1, 0.9)$ and $S^2 \sim \mathcal{N}(0,\Lambda^2)$ where $\Lambda^2 = diag(0.9, 0.1)$. Since $S^1$ and $S^2$ are from different distributions, they are data from two classes. We can see that $S^1$ has the largest variance in one direction and $S^2$ is in the opposite direction. The linear mixing matrix is denoted as $A$, and the observations $Y^1$ and $Y^2$ are the data labeled as class 1 and 2.

Essentially, CSPA seeks the common spatial patterns $\Gamma = A^{-1}$ to filter the observed data $Y^j$, such that the filtered data $\Gamma Y^j$ are uncorrelated in both classes. We generated 200 samples for both classes $S^j$, and $A$ was then randomly generated to form the mixtures $Y^j$ for $j = 1, 2$. Denote $Y = (Y^1, Y^2)$. In the first step, we whiten the data $Y$ via matrix $W$ as in the BSS problem and denote the whitened data as $\widetilde{Y}$. To use the Gibbs sampler, we form the sample covariance matrices $C^1$ and $C^2$ by the samples of $\widetilde{Y}^1$ and $\widetilde{Y}^2$, respectively. We then applied the Gibbs sampler to $C^1$ and $C^2$ to seek an orthogonal matrix $B$ which simultaneously diagonalised the covariance matrices. The filtered data are $Y' = B^T W Y$, which shows that the filtered data of the two classes have largest variances in opposite directions (See Figure 4).

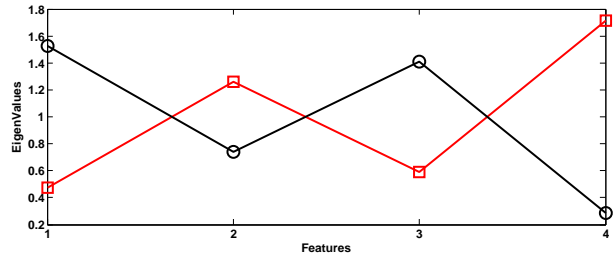

Figure 6. The inferred mean eigenvalues denoted by circles (class 1) and squares (class 2) for the two sample covariance matrices, respectively. Features 1: C3-$\alpha$; 2: C3-$\beta$; 3: C4-$\alpha$; 4: C4-$\beta$.

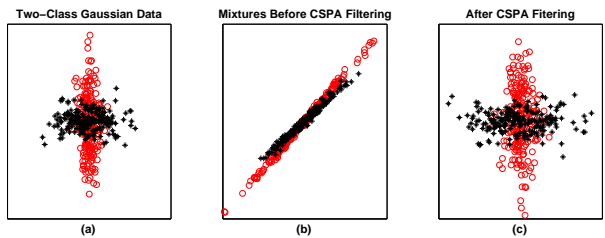

Figure 4. (a) The data from two distributions are labeled as two classes plotted in circle and star, respectively. (b) The linear mixture observations before CSPA filtering. (c) The filtered data indicating that the data of the two classes have largest variances in opposite directions.

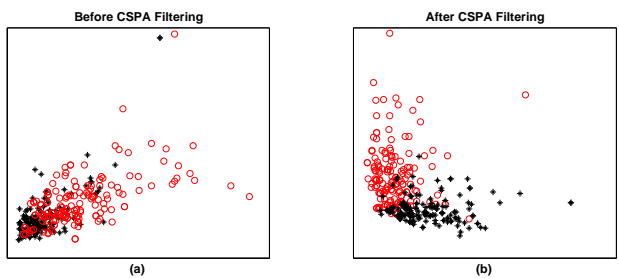

Figure 5. (a) The scatter plots for the features C4-$\alpha$ and C4-$\beta$ before filtering where the circles and stars denote the two classes, respectively. (b) The scatter plots for the features C4-$\alpha$ and C4-$\beta$ of the filtered EEG data.

### 4.4.2. EEG Data

In this section we apply the Gibbs sampler to EEG data in performing CSPA. The data used for this study corresponds to the EEG data set IIIb of the BCI competition III (Blankertz et al., 2006). This data set gathers the EEG signals recorded for three subjects who had to perform motor imagery, i.e. to imagine left or right hand movements. Hence, the two classes to be identified were Left and Right. Before doing the CSPA, we extract features from these EEG signals. We choose to use Band Power (BP) features and finally we have four features for each subject which are denoted as C3-$\alpha$, C3-$\beta$, C4-$\alpha$ and C4-$\beta$. So each data point having four features is labeled as one of the two classes. We do the CSPA for these data points. We denote the two class data as $X^1$ and $X^2$. In the first step we whitened the data by using a whitening matrix $W$ as usual. Then we form the sample covariance matrices $\mathbf{C}_1$ and $\mathbf{C}_2$ for the two classes, respectively. Note that the covariance matrix has size $4 \times 4$. Then the Gibbs sampler was used to seek an orthogonal matrix $B$ to diagonalize the covariance matrices simultaneously. Finally the filtered data of the two classes are then $\widetilde{X}^1 = B^T W X^1$ and $\widetilde{X}^2 = B^T W X^2$. For demonstration, we plotted the scatter plots for the features C4-$\alpha$ and C4-$\beta$ before filtering and after filtering (see Figure 5). From the plots in Figure 5 (a), we see that the data in the two classes have large variance in the same direction, and after the filtering in Figure 5 (b) they have large variances in opposite directions. This can also be demonstrated by the inferred eigenvalues plotted in Figure 6, where the squares denote the eigenvalues for one class and circles for the other. We observe that one class has larger eigenvalues in features C3-$\alpha$ and C4-$\alpha$ and the other class has larger eigenvalues in features C3-$\beta$ and C4-$\beta$. This would be useful for the further classification purpose. It is interesting to note that the sample covariance matrices are positive definite matrices, and thus the inferred eigenvalues finally converged to positive values even though they were initialized in negative values. This could be an advantage for our Gibbs sampler comparing to the algorithms of (Hoff, 2009b) and (Pham, 2001) where the eigenvalues were restricted to positive values.

## 5. Conclusions

We have proposed a Gibbs sampler to simultaneously diagonalize several matrices which are not necessarily symmetric. All the required conditional distributions are known standard distributions. We have compared the Gibbs sampler to some other joint diagonalization algorithms and shown that the Gibbs sampler achieves the state-of-the-art performance on the small examples considered. We also applied the algorithm to BSS, CPCA and CSPS. The Gibbs sampler could be extended to the case where the diagonalizatioin matrix is non-orthogonal.

## Acknowledgments

MG is grateful for financial support from the Engineering and Physical Sciences Research Council (EPSRC) UK, grants EP/J007617/1, EP/E052029/2, EP/H024875/2. MZ is supported by the fundamental research funds for the central universities in China.## References

Amari, S.-I., Cichocki, A., and Yang, H. H. A new learning algorithm for blind source separation. In *NIPS*, 1996.

Belouchrani, A., Meraim, K. Abed, Cardoso, J.-F., and Moulines, E. A blind source separation technique based on second order statistics. *IEEE Trans. Signal Process.*, 45(2):927–949, 1997.

Blankertz, B., Muller, K.R., Krusienski, D.J., Schalk, G., Wolpaw, J.R., Schlogl, A., Pfurtscheller, G., Millan, J.D.R., Schroder, M., and Birbaumer, N. The bci competition iii: Validating alternative approaches to actual bci problems. *IEEE Trans. Neural Syst. Rehab. Eng.*, 14(2):153–159, 2006.

Blankertz, B., Tomioka, R., Lemm, S., M.Kawanabe, and K.-R.Mller. Optimizing spatial filters for robust eeg single-trial analysis. *IEEE Signal Proc. Magazine*, 25(1):41–56, 2008.

Cardoso, J.-F. and Souloumiac, A. Jacobi angles for simultaneous diagonalization. *SIAM J. Mat. Anal. Appl.*, 17(1):161–164, 1996.

Chikuse, Y. *Statistics on special manifolds*. Springer-Verlag, New York, 2003.

Dempster, A. P., Laird, N. M., and Rubin, D. B. maximum likelihood from incomplete data via the em algorithm. *J. R. Statist. Soc. B*, 39(1):1–38, 1977.

Flury, B. Common principal components in k groups. *J. Amer. Statist. Assoc.*, 79(388):892–898, 1984.

Friedland, S. Simultaneous similarity of matrices. *Adv. in Math.*, 50(3):189–265, 1983.

Gelman, A. and Rubin, D. Inference from iterative simulation using multiple sequences. *Statist. Sci.*, 7:457–511, 1992.

Hoff, P.D. Simulation of the matrix Bingham-von Mises-Fisher distribution, with applications to multivariate and relational data. *J. Comput. Graph. Statist.*, 18(2):438–456, 2009a.

Hoff, P.D. A hiearchical eigenmodel for pooled covariance estimation. *J. R. Statist. Soc. B*, 71(5):971–992, 2009b.

Khatri, C. G. and Mardia, K. V. The von mises-fisher matrix distribution in orientation statistics. *J. Roy. Statist. Soc. Ser. B*, 39(1):95–106, 1977.

Koles, Z. J. The quantitative extraction and topographic mapping of the abnormal components in the clinical EEG. *Electroencephalogr. Clin. Neurophysiol.*, 79(6):440–447, 1991.

Kume, A. and Walker, S. G. Sampling from compositional and directional distributions. *Statist. and Comput.*, 16(3):261–265, 2006.

Neal, R. M. Slice sampling. *Ann. Statist.*, 31(3):705–767, 2003.

Pham, D.-T. Joint approximate diagonalization of positive definite matrices. *SIAM J. on Matrix Anal. and Appl.*, 22(4):1136–1152, 2001.

Schwarz, G. Estimating the dimension of a model. *Ann. Statist.*, 6:461–464, 1978.

Souloumiac, A. Nonorthogonal joint diagonalization by combining givens and hyperbolic rotations. *IEEE Trans. Signal Process.*, 57(6):2222–2231, 2009.

van der Veen, A. Joint diagonalization via subspace fitting techniques. In *ICASSP*, pp. 2773–2776, 2001.

Yeredor, A. Non-orthogonal joint diagonalization in the least-squares sense with application in blind source separation. *IEEE Trans. on Sig. Proc.*, 50(7):1545–1553, 2002.

Zhong, M., Girolami, M., Faulds, K., and Graham, D. Bayesian methods to detect dye-labelled dna oligonucleotides in multiplexed raman spectra. *J. R. Statist. Soc. C*, 60(2):187–206, 2011.

Ziehe, A., Laskov, P., Nolte, G., and Muller, K-R. A fast algorithm for joint diagonalization with non-orthogonal transformations and its application to blind source separation. *JMLR*, 5:777–800, 2004.